\documentclass[runningheads]{llncs}
\usepackage[T1]{fontenc}
\usepackage{graphicx}
\usepackage{tcolorbox}
\usepackage{amsmath}
\usepackage{amssymb}
\usepackage{booktabs}
\usepackage{multirow}
\usepackage{array} 
\usepackage{arydshln}
\usepackage{subfig}
\usepackage{hyperref}

\begin{document}
\title{MMSRARec:  Summarization and Retrieval Augumented Sequential Recommendation Based on Multimodal Large Language Model}
\titlerunning{MMSRARec}
%
\author{Haoyu Wang\inst{1} \and
Yitong Wang\inst{1} \and
Jining Wang\inst{1}}
%
%
\institute{ College of Computer Science and Artificial Intelligence, Fudan University
\\
\email{wanghy24@m.fudan.edu.cn}}
\maketitle              
\begin{abstract}
Recent advancements in Multimodal Large Language Models (MLLMs) have demonstrated significant potential in recommendation systems. However, the effective application of MLLMs to multimodal sequential recommendation remains unexplored: \textit{A)} Existing methods primarily leverage the multimodal semantic understanding capabilities of pre-trained MLLMs to generate item embeddings or semantic IDs, thereby enhancing traditional recommendation models. These approaches generate item representations that exhibit limited interpretability, and pose challenges when transferring to language model-based recommendation systems. \textit{B)} Other approaches convert user behavior sequence into image-text pairs and perform recommendation through multiple MLLM inference, incurring prohibitive computational and time costs. \textit{C)} Current MLLM-based recommendation systems generally neglect the integration of collaborative signals. To address these limitations while balancing recommendation performance, interpretability, and computational cost, this paper proposes \textbf{M}ulti\textbf{M}odal \textbf{S}ummarization-and-\textbf{R}etrieval-\textbf{A}ugmented Sequential \textbf{Rec}ommendation (MMSRARec). Specifically, we first employ MLLM to summarize items into concise keywords and fine-tune the model using rewards that incorporate summary length, information loss, and reconstruction difficulty, thereby enabling adaptive adjustment of the summarization policy. Inspired by retrieval-augmented generation, we then transform collaborative signals into corresponding keywords and integrate them as supplementary context. Finally, we apply supervised fine-tuning with multi-task learning to align the MLLM with the multimodal sequential recommendation. Extensive evaluations on common recommendation datasets demonstrate the effectiveness of MMSRARec, showcasing its capability to efficiently and interpretably understand user behavior histories and item information for accurate recommendations.

\end{abstract}
\section{Introduction}
\begin{figure*}[htbp]
\centering
\includegraphics[width=\linewidth]{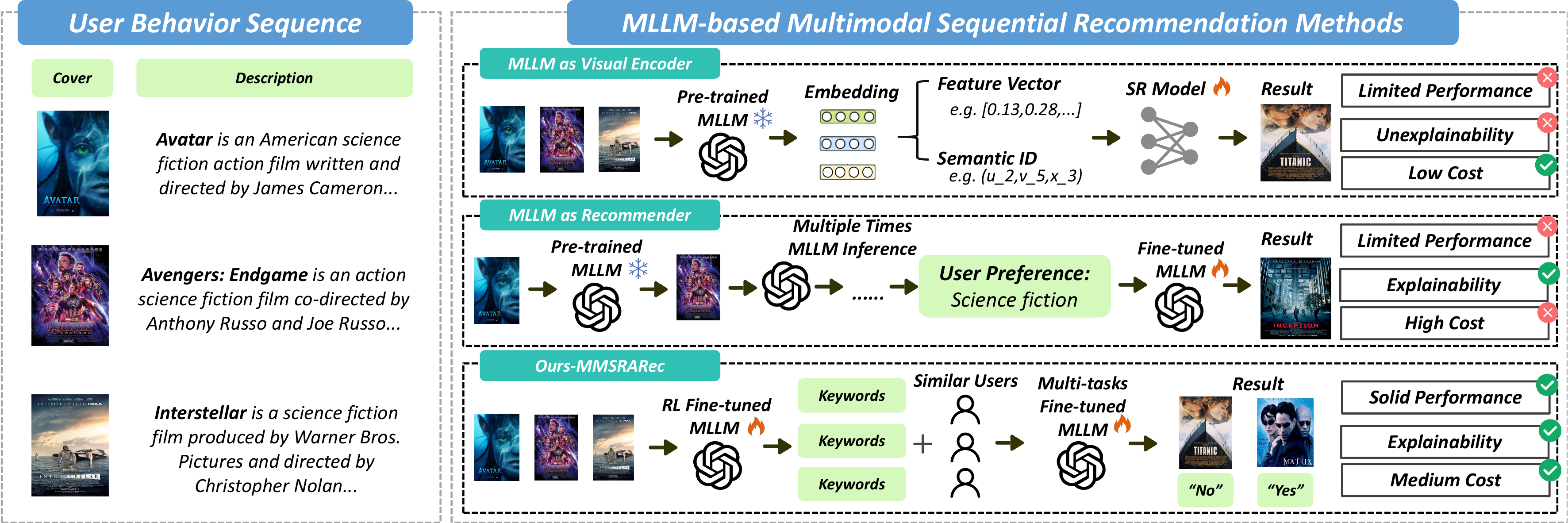} 
\caption{Comparison between existing multimodal sequential recommendation methods based on MLLM and our proposed method.}
\label{fig:fig1}
\end{figure*}

Recommendation Systems (RSs) \cite{resnick1997recommender} serve as tools in various online applications to help users filter out irrelevant information and discover items of interest.    With the advancement of the Internet, multimodal item information such as covers, detailed images, and textual descriptions has become a crucial input source for RSs.    Among these systems, Sequential Recommendation (SR) \cite{boka2024survey} methods have gained prominence by capturing dynamic user interests.    These methods encode users and items as unique identifiers, utilize historical interaction data to learn sequential behavior patterns, and incorporate multimodal information as supplementary input to enhance recommendations.

Recent advances in Multimodal Large Language Models (MLLMs) have demonstrated significant potential in RSs \cite{zhou2025large}. Several approaches have been explored to adapt MLLMs for multimodal SR tasks, which can be broadly categorized into two main directions, as illustrated in Fig. \ref{fig:fig1}. The first approach treats pre-trained MLLMs as visual encoders, where multimodal information is fed into the MLLM at an early stage to obtain item representations \cite{zhang2025notellm,dang2025mllmrec,luo2024molar}. These representations are then used either directly as feature embeddings or further processed into semantic IDs, which are subsequently integrated into traditional multimodal recommendation models. However, such methods suffer from several limitations. First, they leverage only the pre-trained MLLMs, overlooking the distribution shift between MLLM pre-training corpora and real-world recommendation data. Second, the generated item representations are non-interpretable vector embeddings rather than readable natural language, which are inherently incompatible with the textual space of language models. Consequently, they fail to directly exploit the powerful generalization and reasoning capabilities inherent in the language models themselves.

The second approach reformulates sequential recommendation tasks as Natural Language Processing (NLP) tasks \cite{ye2025harnessing}.    In this paradigm, item information from the user's interaction history is directly input into the MLLM, which then performs multiple inferences to generate recommendations.    This methodology capitalizes on the MLLM's semantic understanding and in-context learning capabilities, offering better interpretability.    Nevertheless, the computational and temporal costs for multiple times MLLM inferences renders it impractical for online deployment.    Besides, these methods neglect collaborative signals, causing the models to overemphasize item features while overlooking latent user behavior patterns.    These limitations and challenges indicate that the effective application of MLLMs in multimodal SR remains largely unexplored.

To address these challenges and holistically balance recommendation performance, interpretability, and cost, this paper proposes \textbf{M}ulti\textbf{M}odal \textbf{S}ummarization-and-\textbf{R}etrieval-\textbf{A}ugmented Sequential \textbf{Rec}ommendation (MMSRARec).   Our recommendation method consists of three stages, as illustrated in Fig. \ref{fig:model}.    Specifically, to align multimodal SR with NLP tasks while avoiding multiple online inferences, we first devise multimodal summarization stage.   In this stage, multimodal item information is summarized into representative sets of keywords using MLLM.   The summarization strategy is adaptively optimized via Reinforcement Learning with Verifiable Rewards (RLVR) \cite{mroueh2025reinforcement}, which jointly considers summary length, information loss, and reconstruction difficulty.   This stage can be performed offline, and its outputs are interpretable natural language keywords rather than unreadable vectors or semantic IDs.

Second, to incorporate collaborative signals into the recommendation framework, we introduce similar-user retrieval stage.   Based on traditional item ID features, users with similar historical behaviors are retrieved, and their subsequent interactions are converted into corresponding keywords and  are then integrated into the input as contextual information.  Finally, we design four types of recommendation tasks and train the MLLM via Parameter-Efficient  Fine-Tuning (PEFT) \cite{ding2023parameter} in  multi-task learning stage.   This enables the model to understand and adapt to multimodal sequential recommendation tasks while mitigating overfitting.  Extensive evaluations on three commonly-used recommendation datasets demonstrate the effectiveness of MMSRARec, confirming its capability to comprehend user behavior history and item information while achieving efficient and interpretable recommendations.

Our contributions can be summarized as follows:

\begin{itemize}
    \item We propose MMSRARec, a novel approach for sequential recommendation that leverages multimodal large language models. By summarizing item information, retrieving similar users, and aligning the SR task with multi-task learning, MMSRARec effectively utilizes the semantic understanding and in-context learning capabilities of MLLMs to achieve accurate and interpretable recommendations.
    \item To the best of our knowledge, our work is the first attempt that introduces item summarization into natural language keywords and employs reinforcement learning to fine-tune MLLMs, while incorporating collaborative signals into the language model via retrieval of similar users. These innovations address key challenges in existing MLLM-based methods and enhance recommendation performance.
    \item We conduct sufficient experiments on three real-world recommendation datasets, demonstrating that MMSRARec effectively enhances recommendation performance, interpretability, and inference efficiency for multimodal sequential recommendation.
\end{itemize}

\section{Related Work}
\subsection{Multimodal Large Language Model} \label{RW:MLLM}
Multimodal Large Language Models (MLLMs) based on multimodal pre-training \cite{zhu2024multimodal} have advanced rapidly in recent years, achieving remarkable performance across a variety of downstream vision-language tasks such as visual question answering, grounding , and image captioning.    With the progress of visual instruction tuning, state-of-the-art MLLMs including GPT-4o \cite{hurst2024gpt}, Gemini \cite{team2023gemini}, and Qwen-VL \cite{bai2025qwen2} are capable of effectively understanding human intentions and visual inputs, as well as performing complex multimodal in-context learning in response to instructions.    However, directly applying MLLMs to sequential recommendation tasks remains challenging.    Firstly, due to constraints on the model’s context length, it is difficult to input complete user behavior sequences into MLLMs.    Secondly, MLLMs still exhibit limited capability in comprehending multiple images, which hinders their ability to interpret temporal trends in user interactions.    Lastly, compared to traditional sequential recommendation models, MLLMs require significantly more time and computational resources for inference.    Our proposed MMSRARec overcomes the limitations of context length and the high cost of multi-image reasoning through keyword-based compression, making it more suitable for real-world recommendation scenarios.

\subsection{MLLM-Based Multimodal Sequetial Recommendation}
Application of MLLMs to multimodal sequential recommendation tasks primarily follows two main directions.    The first approach leverages pre-trained MLLMs as feature encoders.    For example, NoteLLM-2 \cite{zhang2025notellm} employs a late fusion mechanism to directly integrate visual information with textual data, representing both image and text content as learnable tokens.    Molar \cite{luo2024molar} integrates multiple content modalities with ID information, using MLLMs to generate unified item representations from both textual and non-textual data.    MLLMRec \cite{dang2025mllmrec} utilizes MLLMs to convert item images into high-quality semantic descriptions, which are then fused with the item's textual metadata.    However, such methods primarily exploit the representational capacity of pre-trained MLLMs while overlooking the distribution shift between MLLM training data and real-world recommendation system data, and they often lack interpretability.    The second approach formulates multimodal sequential recommendation as a natural language processing task, with MLLMs directly serving as the recommender.    For instance, MLLM-MSR \cite{ye2025harnessing} summarizes user preferences into textual form through multiple inferences with large language model, after which the MLLM makes recommendations by combining these summarized preferences with multimodal item information.    Nevertheless, the multiple inference steps required by this paradigm entail substantial computational and time costs.    In contrast, MMSRARec employs RLVR to guide the MLLM in adaptively adjusting its representation strategy based on recommendation data, enabling effecient recommendations through once inference.

\section{Method}

\subsection{Problem Formulation}
We formulate the multimodal sequential recommendation task as follows: given a user $u \in \mathcal{U}$ and a chronologically ordered sequence of historically interacted items $\mathcal{H}_u=\{\mathcal{I}_1,\mathcal{I}_2,...,\mathcal{I}_n\}$, where each item $\mathcal{I}_i$ is represented by its ID, image, and textual description as $\mathcal{I}_i=(id_i, img_i, text_i)$, the objective is to predict the item $\mathcal{I}_{n+1}$ that the user is most likely to interact with at the $n+1$-th time step.

\subsection{Overview}
\begin{figure*}[t]
\centering
\includegraphics[width=\linewidth]{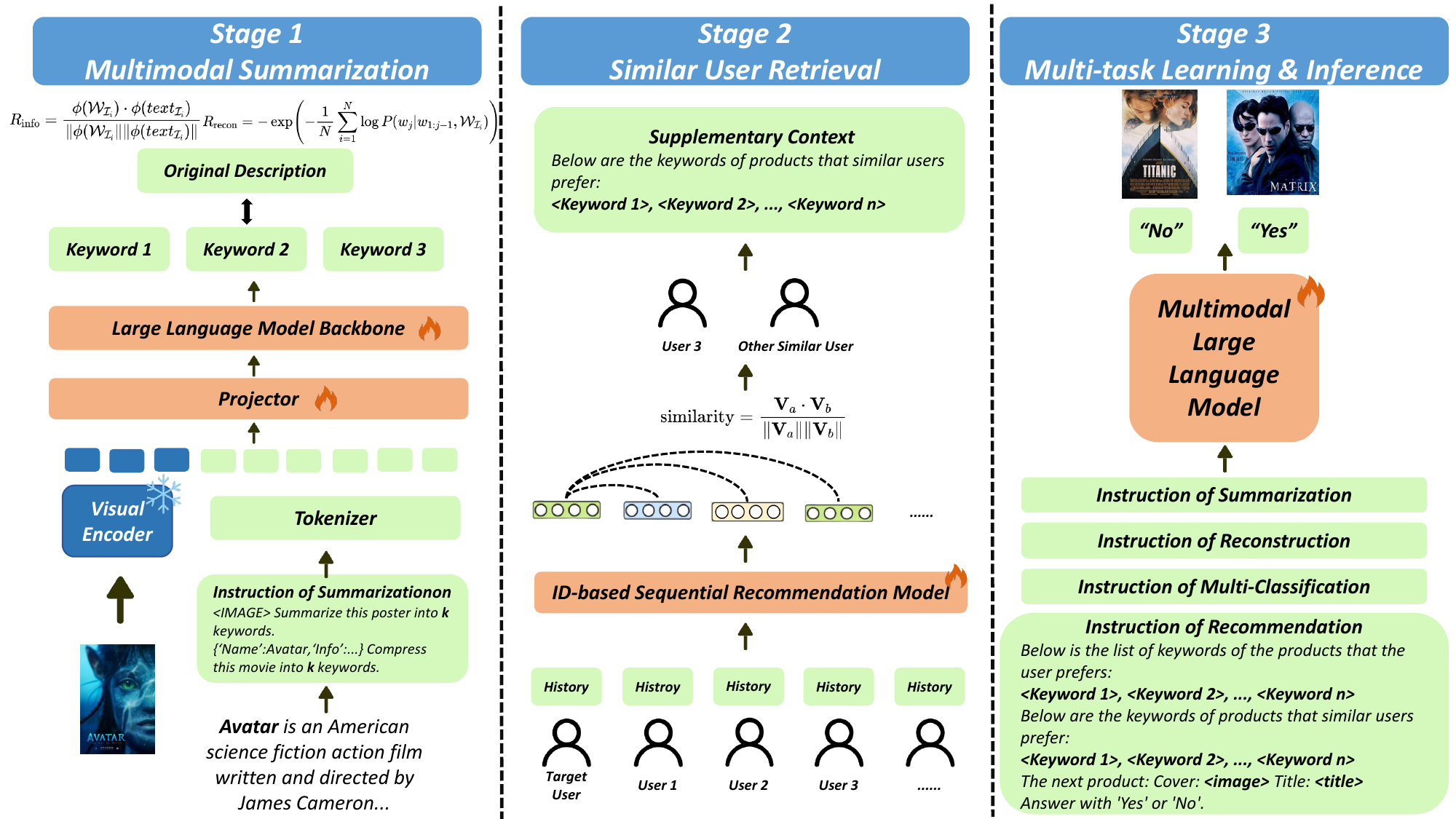} 
\caption{The framework of MMSRARec.}
\label{fig:model}
\end{figure*}

Balancing recommendation performance, interpretability, and cost, we propose a three-stage MLLM-based multimodal sequential recommendation pipeline, termed MMSRARec, as illustrated in Fig. \ref{fig:model}. First, to compress multimodal user interaction information into a context length acceptable to the MLLM, during the multimodal summarization stage, we summarize the user behavior history $\mathcal{H}_u$ into a set of natural language keywords $\mathcal{K}_u=\{\mathcal{W}_{\mathcal{I}_1},\mathcal{W}_{\mathcal{I}_2},...,\mathcal{W}_{\mathcal{I}_n}\}$. Subsequently, to introduce collaborative signals into the MLLM, we retrieve users with similar interaction histories $\mathcal{S}_u=\{u',u'',...\}$ based on user ID information—which is typically challenging for MLLMs to leverage directly—and incorporate the keywords $\{\mathcal{K}_{u^*}\}_{u^* \in \mathcal{S}_u}$ corresponding to the subsequent interactions of these similar users as auxiliary contextual information. Finally, we frame the multimodal sequential recommendation task as a natural language processing task by using  $(K_u, \{\mathcal{K}_{u^*}\}_{u^* \in \mathcal{S}_u},id_{n+1},img_{n+1},text_{n+1})$ as prompts, and perform parameter-efficient fine-tuning of the MLLM driven by multi-task learning.

\subsection{Multimodal Summarization Stage}

As discussed in Sec. \ref{RW:MLLM}, attempts to apply MLLMs to SR are constrained by the fact that the number of images associated with a user’s historical interactions often exceeds the MLLM’s context window.  Consequently, most approaches resort to using the MLLM solely as feature encoder.  However, this practice merely leverages the pretrained MLLM and overlooks the mismatch between the MLLM’s training data and real-world recommendation data.  Moreover, an MLLM-based encoder represents item information as uninterpretable feature vectors, which are intrinsically incompatible with the textual space of language models, thereby necessitating additional alignment training.

\begin{table*}
\vspace{-1em}
\resizebox{\linewidth}{!}{
\begin{tcolorbox}
\small
You are an expert in recommendation. Below is an Amazon product:

Cover: <image>

Title: <title>

Description: <description>

Please summarize the cover and content of this product with several keywords respectively. The summary should be concise and accurate. Output using the following template:

Cover: <image keyword 1>,<image keyword 2> ...

Content: <content keyword 1>,<content keyword 2> ...
\end{tcolorbox}}
\caption{Prompt template for multimodal summarization.}
\label{tab:prompt_summarization}
\vspace{-2em}
\end{table*}

Recent research on text-only LLMs has shown that extracting item attributes as specific natural-language keywords via an LLM-style \textbf{\textit{summarization}} procedure can also benefit recommendation \cite{zhang2025llm,chen2024hllm}.  Inspired by this, we design a Multimodal Summarization Stage to represent multimodal item information as interpretable keywords that serve as context for language-model-based recommendation.  Specifically, for an item $\mathcal{I}_i$, we prompt the MLLM with the template shown in Table \ref{tab:prompt_summarization} to obtain keywords $\mathcal{W}_{\mathcal{I}_i}$.  To mitigate modality bias, where an MLLM may ignore information from a particular modality, we ask the model to summarize $img_i$ and $text_i$ separately, thereby reducing item-information loss during summarization.

Furthermore, to adapt MLLMs to recommendation data, we employ Reinforcement Learning with Verifiable Rewards (RLVR) \cite{mroueh2025reinforcement} to stimulate the summarization ability of MLLMs.  We design three distinct rewards to train the MLLM to adaptively adjust its summarization policy. Information Reward computes the semantic similarity between the model’s summary and the original item description, encouraging the model to preserve key information about the item.  Reconstruction Reward measures the perplexity of reconstructing the original description from keywords generated by the MLLM, guiding the model to use interpretable and easily understandable keywords.  Length Reward penalizes the number of keywords in the summary, preventing the model from reward hacking by generating an excessive number of words.  The specific formulations are as follows:

\begin{align}
    R_{\text{info}} &= \frac{ \phi(\mathcal{W}_{\mathcal{I}_i}) \cdot \phi(text_{\mathcal{I}_i}) }{ \| \phi(\mathcal{W}_{\mathcal{I}_i}\| \|\phi(text_{\mathcal{I}_i})\| } \\
    R_{\text{recon}}&= - \exp\left( -\frac{1}{N} \sum_{i=1}^{N} \log P(w_j | w_{1:j-1}, \mathcal{W}_{\mathcal{I}_i}) \right) \\
    R_{\text{len}} &= - |\mathcal{W}_{\mathcal{I}_i}| \\
    R&=\alpha R_{\text{info}} + \beta R_{\text{recon}} + \gamma R_{\text{len}}
\end{align}

We adpot the Group Relative Policy Optimization (GRPO) \cite{shao2024deepseekmath} algorithm for reinforcement learning, which eliminates the need for a specifice critic network by leveraging group-level reward statistics. In summary, the multimodal summarization stage maps multimodal item information into a natural language space by applying RLVR on recommendation data to the MLLM.  This process effectively summarizes item information while preserving interpretability and the integrity of item feature.  This stage can be conducted offline, where items are pre-processed into keywords, thereby reducing the overhead of online inference.


\subsection{Similar-User Retrieval Stage}

Another significant challenge in applying MLLMs to SR lies in how to incorporate collaborative information into the MLLM. Given the finite context window of MLLMs, it is computationally prohibitive to feed the entire interaction history of all users into the model. Inspired by the Retrieval-Augmented Generation (RAG) \cite{lewis2020retrieval} paradigm, we introduce item ID information, which is inherently difficult for MLLMs to utilize directly.

First, we train a conventional SR model (e.g., SASRec \cite{kang2018self}) using item IDs. Then the interaction histories of all users are fed into the pre-trained model to obtain feature embedding $\{e_u\}_{u \in \mathcal{U}}$ representing historical behaviors. Subsequently, for each target user $u$, we retrieve $k$ similar users $\mathcal{S}_u=\{u',u'',...\}$ based on cosine similarity $sim(e_{u'},e_u)$. Crucially, we ensure that all retrieved similar users are sourced from the knowledge base (i.e., the training set), so their subsequent interacted items are known. Finally, we aggregate the keywords $\{\mathcal{K}_{u^*}\}_{u^* \in \mathcal{S}_u}$ corresponding to the items subsequently interacted with by these similar users, which serve as auxiliary contextual information for the MLLM in the next stage of recommendation.

\subsection{Multi-Task Learning Stage}

\begin{table}[ht]
\vspace{-1em}
\resizebox{\linewidth}{!}{
\begin{tcolorbox}
\small

\textbf{Basic Instruction}

You are an expert in recommendation. Below is the list of keywords of the products that the user prefers:

<Keyword 1>, <Keyword 2>, ..., <Keyword n>

Below are the keywords of products that relevant users prefer:

<Keyword 1>, <Keyword 2>, ..., <Keyword n>

The next product: Cover: <image> Title: <title> Description: <description>

Based on the above information, please predict whether the user is likely to purchase this product. Answer with 'Yes' or 'No'. \\

\textbf{Instruction of Multi-Classification}

You are an expert in recommendation. Below is the list of keywords of the products that the user prefers:

<Keyword 1>, <Keyword 2>, ..., <Keyword n>

Below are the keywords of products that relevant users prefer:

<Keyword 1>, <Keyword 2>, ..., <Keyword n>

Candidate 1: Cover: <image> Title: <title> Description: <description>

Candidate 2: Cover: <image> Title: <title> Description: <description>

... Based on the above information, please predict which candidate products the user will purchase. Answer with the serial number of this product. \\

\textbf{Instruction of Reconstruction}

You are an expert in recommendation.Below are the keywords of a product:

<Keyword 1>, <Keyword 2>, ..., <Keyword n>

Based on these keywords, please describe this product in its entirety. \\

\textbf{Instruction of Summarization}

(same as Table 1) ...
\end{tcolorbox}}
\caption{Prompt template for multimodal sequential recommendation.}
\label{tab:prompt_recommendation}
\vspace{-3em}
\end{table}

After completing the preparations in the summarization and retrieval stages, the contextual information required by the MLLM for recommendation has been fully integrated. According to the definition of SR, given user interaction history $\mathcal{H}_u$, an MLLM-based recommendation system leverages prompts containing keywords of interacted items $\mathcal{K}_u$, keywords of items interacted by similar users $\{\mathcal{K}_{u^*}\}_{u^* \in \mathcal{S}_u}$, textual descriptions $text_{n+1}$ and image $img_{n+1}$ of candidate item, as well as designed recommendation instructions, to predict the probability of user interaction with candidate items. During the recommendation process, we employ a prompt template as illustrated in Table \ref{tab:prompt_recommendation} and compute the probability of recommending candidate items based on the probability distribution of the first token generated by the MLLM:
$
p=\frac{p('\text{yes}')}{p('\text{yes}')+p('\text{no}')}
$

To align the capabilities of MLLMs with SR, the model must be fine-tuned to minimize the discrepancy between predicted and actual user interactions.  We construct the fine-tuning data using a combination of positive and negative sampling: positive samples represent items with which the user interacts in the future, while negative samples are randomly selected from items that the user has not interacted with.  This approach enables the model to distinguish between relevant and irrelevant items through contrastive learning, thereby improving its predictive accuracy.  Furthermore, inspired by LC-Rec \cite{zheng2024adapting}, we uniformly construct multiple types of recommendation instructions along with their corresponding training data.  This strategy helps the model develop a comprehensive understanding of the recommendation task while mitigating overfitting, as illustrated Table \ref{tab:prompt_recommendation}.

Fine-tuning employs the next-token prediction paradigm, training the model to predict subsequent tokens in a sequence based on preceding tokens. This ensures the generation of coherent and contextually relevant outputs from the input sequence. The supervised fine-tuning loss function is defined as:
$$
\mathcal{L}_{\text{SFT}} = -\frac{1}{N} \sum_{i=1}^{N} \sum_{t=1}^{T} \log P(y_t^{(i)} | x^{(i)}, y_{<t}^{(i)})
$$
Furthermore, we adopt Low-Rank Adaption (LoRA) \cite{hu2022lora} following the parameter-efficient fine-tuning (PEFT) \cite{ding2023parameter} framework, thereby accelerating the training process while preserving the model's inherent in-context learning capability.

\section{Experiment}
\subsection{Dataset}

We utilized open-source and real-world datasets from diverse recommendation domains to ensure broad applicability and robust validation: (1) \textbf{Microlens} \cite{ni2023content}, a micro-video recommendation dataset; (2) \textbf{Amazon Baby} \cite{he2016ups}, from the e-commerce domain, representing dense purchasing behavior in the baby product category; and (3) \textbf{Amazon Games} \cite{mcauley2015image}, which reflects user preferences in the digital goods sector. All datasets comprise user-item interactions, product descriptions, and images.   During the preprocessing stage, we filtered out users and items with fewer interactions to ensure that user historical behavior sequences met a minimum length threshold.   Following the practice of MLLM-MSR \cite{ye2025harnessing}, for each user $u$ with a historical sequence of length $n$, we treated the user's interaction after time step $n+1$ as positive samples.   Additionally, we randomly selected 20 items with which the user had not interacted as negative samples. We randomly split all impression item lists into training, validation, and test sets in an 8:1:1 ratio.  Detailed statistics for these datasets are provided in Table \ref{tab:dataset}.

\begin{table}[ht]
\vspace{-3em}
\centering
\caption{The Statistics of Datasets}
\label{tab:dataset}
\begin{tabular}{lccc}
\hline
\textbf{Dataset} & \textbf{Microlens} & \textbf{Amazon Baby} & \textbf{Amazon Game} \\
\hline
\#User         & 25411     & 41081      & 38808      \\
\#Item         & 20276     & 14393      & 13379      \\
\#Interaction  & 223263    & 400876     & 352136     \\
\#Avg Seqlen   & 11.35     & 13.65      & 13.23      \\
\hline
\end{tabular}
\vspace{-3em}
\end{table}

\subsection{Experimental Setup}
To assess the performance of baseline and our proposed method for multimodal sequential recommendations, we utilize HR@5, NDCG@5 and AUC as evaluation metrics.  All models are evaluated in Python 3.10 using PyTorch \cite{paszke2019pytorch}. All experiments are carried out on a workstation equipped with 8×NVIDIA A800 GPUs running Ubuntu 24.04.2 LTS, using PyTorch 2.6.0 with CUDA 12.9.  In the experiments, we conduct inference and training of open-source models based on the MS-Swift \cite{zhao2025swift} framework. All experiments are repeated three times under the same random seed to compute average value. Besides the parameter analysis experiments, in all other experiments, we set the length of user behavior sequences to 5 and the number of retrieved similar users to 3. 

\subsection{Baseline}
To evaluate the effectiveness of our proposed method, we select several mainstream methods for comparison covering four categories of recommendation systems:

\begin{enumerate}
    \item Basic SR Models: These models utilize only item IDs and collaborative information for recommendation.     SASRec \cite{kang2018self} employs self-attention mechanisms to capture long-term dependencies, BERT4Rec \cite{sun2019bert4rec} adopts bidirectional self-attention to model user behavior sequences.
    \item Multimodal SR Models: Beyond item IDs, these models leverage visual information from item images.      MMSR \cite{hu2023adaptive} achieves adaptive fusion of multimodal features via graph structures.      HM4SR \cite{zhang2025hierarchical} introduces a two-level Mixture-of-Experts (MoE) architecture combined with a multi-task learning strategy to capture dynamic user interests.
    \item LLM-based SR Models: These models harness the semantic understanding and in-context learning capabilities of LLMs for recommendation, using both item IDs and textual information.      HLLM \cite{chen2024hllm} adopts a two-tower architecture: the first LLM layer extracts content features from item textual descriptions, and the second LLM predicts future user interests based on interaction history.      LLM-ESR \cite{liu2024llm} enhances traditional SR models by incorporating semantic embeddings generated by LLMs.
    \item MLLM-based SR Models: These models comprehensively utilize multimodal information and leverage the visual comprehension capabilities of MLLMs for recommendation.      MLLM-MSR \cite{ye2025harnessing} converts item images and text into natural language descriptions via MLLMs and infers user preferences through multiple rounds of LLM reasoning.      MLLMRec \cite{dang2025mllmrec} employs MLLMs to transform item information into high-quality semantic descriptions, which are then integrated into an item-item graph learning framework for recommendation.
\end{enumerate}

\subsection{Main Results}

\begin{table}[t]
\centering
\caption{Comparison of sequential recommendation performance of different models. The optimal results are marked in \textbf{bold}, and the suboptimal results are marked with \underline{underlines}.}
\resizebox{\textwidth}{!}{
\begin{tabular}{l|ccc|ccc|ccc}
\hline
Model & \multicolumn{3}{c|}{\textbf{Microlens}} & \multicolumn{3}{c|}{\textbf{Amazon Baby}} & \multicolumn{3}{c}{\textbf{Amazon Games}} \\
 & HR@5 & AUC & NDCG@5 & HR@5 & AUC & NDCG@5 & HR@5 & AUC & NDCG@5 \\
\hline
SASRec (ICDM'18)      & 66.64 & 74.02 & 31.20 & 58.24 & 71.06 & 27.56 & 62.36 & 80.54 & 29.66 \\
BERT4Rec (CIKM'19)    & 54.47 & 72.55 & 28.99 & 50.85 & 73.45 & 25.41 & 55.15 & 73.68 & 27.25 \\
MMSR (CIKM'23)        & 69.85 & 78.87 & 49.75 & 65.44 & 79.01 & 44.22 & 68.93 & 81.19 & 47.53 \\
HM4SR (WWW'25)        & 76.46 & 80.23 & 57.86 & 72.37 & 82.15 & 53.60 & 75.27 & 83.59 & 55.81 \\
HLLM ('24)            & 71.23 & 76.54 & 30.52 & 67.86 & 77.28 & 26.98 & 70.11 & 78.84 & 28.49 \\
LLM-ESR (NIPS'24)     & 68.25 & 77.81 & 62.67 & 64.59 & 78.63 & 55.85 & 67.20 & 79.45 & 57.92 \\
MLLM-MSR (AAAI'25)    & 77.42 & 83.17 & 54.26 & 73.92 & \underline{84.39} & 58.57 & 76.39 & \underline{85.69} & \textbf{63.73} \\
MLLMRec ('25)         & \underline{81.32} & \underline{83.25} & \textbf{68.25} & \underline{77.61} & 81.79 & \underline{61.24} & \underline{79.86} & 82.16 & 58.40 \\
\hline
\textbf{MMSRARec}         & \textbf{85.09} & \textbf{84.36} & \underline{66.12} & \textbf{81.50} & \textbf{85.27} & \textbf{62.82} & \textbf{83.74} & \textbf{85.81} & \underline{63.14} \\
\hline
\end{tabular}
}
\label{tab:mainresult}
\end{table}

Table \ref{tab:mainresult} presents a comparison between our proposed MMSRARec and  baseline methods across three real-world datasets.  MMSRARec achieves the best or second-best performance across all evaluation metrics.  Specifically, it attains Hit Rate @5 scores of 85.1, 81.5, and 83.7, significantly outperforming the suboptimal models, which achieve 81.3, 77.6, and 79.8, respectively.  These results indicate that MMSRARec is capable of capturing user preferences and recommending appropriate items accordingly.  Overall, recommendation methods leveraging multimodal information demonstrate superior performance compared to those relying solely on ID and textual data.  Moreover, approaches based on LLM or MLLM outperform traditional deep neural network-based models.  This underscores the importance of integrating MLLMs to interpret multimodal item information within recommendation systems.

Another advantage of language model-based recommendation systems lies in their interpretability.  Tables \ref{tab:case}present recommendation cases from the Microlens and Amazon Baby datasets.  The keywords summarized during the multimodal summarization stage effectively capture the key characteristics of items, significantly reduce the required input context length, and offer stronger interpretability compared to non-readable vectors or semantic IDs.  For instance, in Microlens, the keywords generated by MMSRARec indicate that the user’s historical videos are all related to the mobile game "Kings of Glory."  As a result, the model recommends videos related to this game while excluding those associated with another game, "Genshin."  These examples demonstrate that leveraging MLLMs to summarize keywords can effectively extract essential item information and align MLLM with the space of recommendation, thereby enhancing both recommendation accuracy and interpretability.  In contrast, existing methods that rely solely on pretrained MLLMs for feature extraction exhibit certain limitations.

\begin{table}[ht]
\centering
\caption{Recommended cases on Microlens and Amazon Baby dataset.}
\label{tab:case}
\resizebox{\textwidth}{!}{
\begin{tabular}{>{\raggedright\arraybackslash}m{2cm}|>{\centering\arraybackslash}m{2cm}|>{\centering\arraybackslash}m{2cm}|>{\centering\arraybackslash}m{2cm}|>{\centering\arraybackslash}m{2cm}|>{\centering\arraybackslash}m{2cm}::>
{\centering\arraybackslash}m{2cm}|>{\centering\arraybackslash}m{2cm}}
\hline
\textbf{User Behavior} & \textbf{1} & \textbf{2} & \textbf{3} & \textbf{4} & \textbf{5} & \textbf{Candidate} & \textbf{Candidate} \\
\hline
\textbf{ID} & 2259 & 6810 & 3325 & 12352 & 17383 & 8710 & 6398 \\
\hline
\textbf{Cover} & 
\includegraphics[width=1.5cm, height=1.5cm, keepaspectratio]{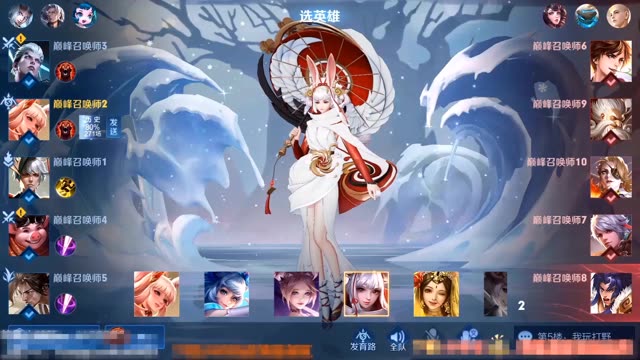} &
\includegraphics[width=1.5cm, height=1.5cm, keepaspectratio]{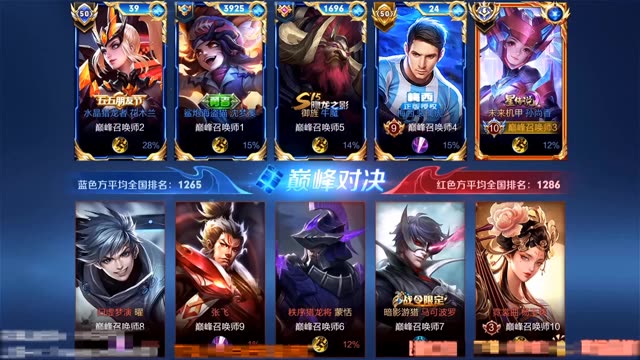} &
\includegraphics[width=1.5cm, height=1.5cm, keepaspectratio]{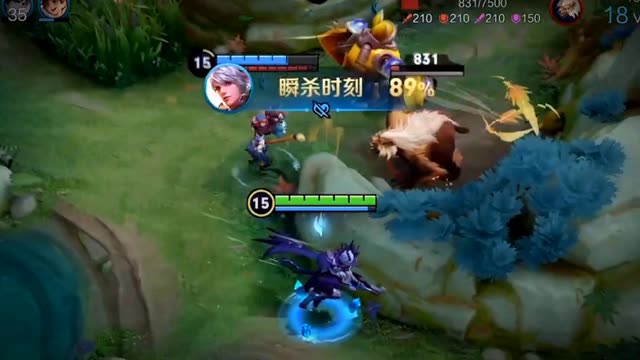} &
\includegraphics[width=1.5cm, height=1.5cm, keepaspectratio]{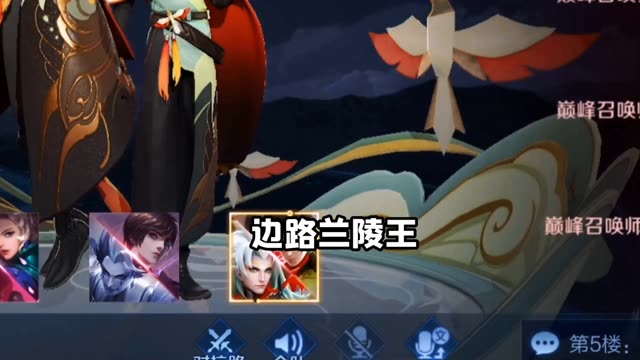} &
\includegraphics[width=1.5cm, height=1.5cm, keepaspectratio]{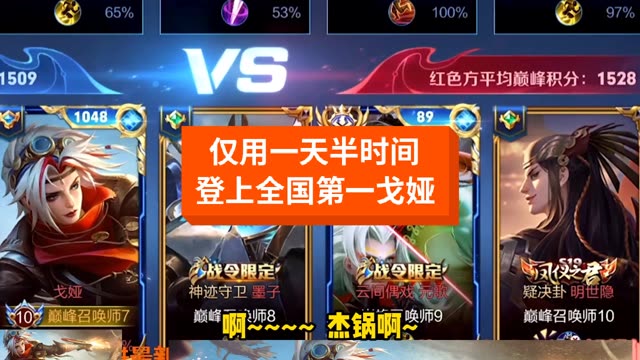} &
\includegraphics[width=1.5cm, height=1.5cm, keepaspectratio]{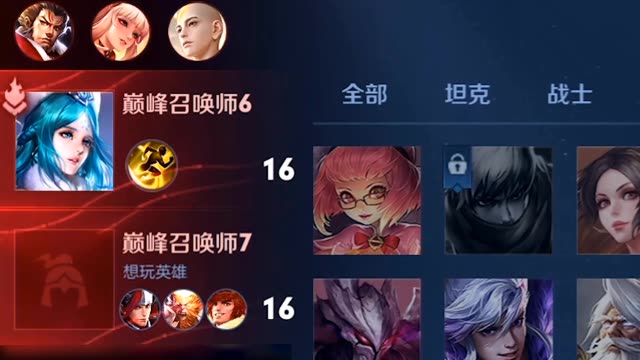} &
\includegraphics[width=1.5cm, height=1.5cm, keepaspectratio]{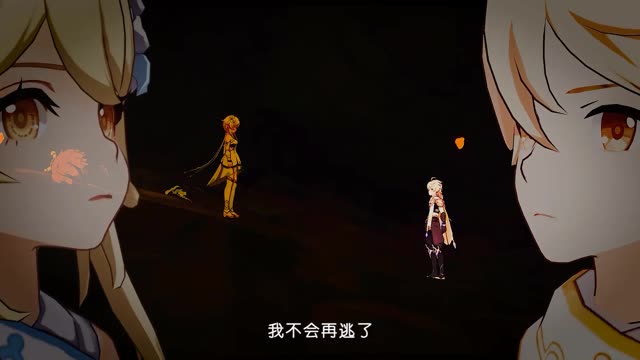} \\
\hline
\textbf{Title} & Gong sun li & National best shooter & Solo & Hades Side Road & New hero Goya & Shaoxing and Yaoyao & Not enemy, just different positions \\
\hline
\textbf{Keywords by MMSRARec} & character depiction, Kings of Glory, Gongsun Li, game strategies & mobile game, character depiction, game play experience  &  mobile game character, solo kill, Kings of Glory & character depiction, Hades Side Road, game play & Kings of Glory, Goya, character depiction, player achieves & mobile game, \textcolor{red}{Kings of Glory}, strategic moment & \textcolor{blue}{Genshin character}, Dawn Roses, Challenge \\
\hline
\textbf{Recommend} & - & - & - & - & - & \checkmark & $\times$ \\
\hline
\\
\hline
\textbf{ID} &457 & 1178 & 5056 & 1022 & 1177 & 134 & 13420 \\
\hline
\textbf{Cover} & 
\includegraphics[width=1.5cm, height=2cm, keepaspectratio]{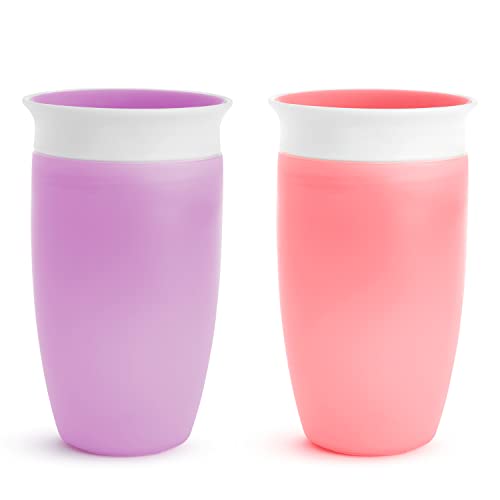} &
\includegraphics[width=1.5cm, height=2cm, keepaspectratio]{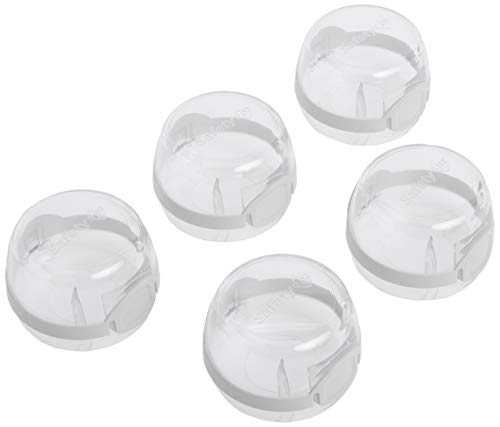} &
\includegraphics[width=1.5cm, height=2cm, keepaspectratio]{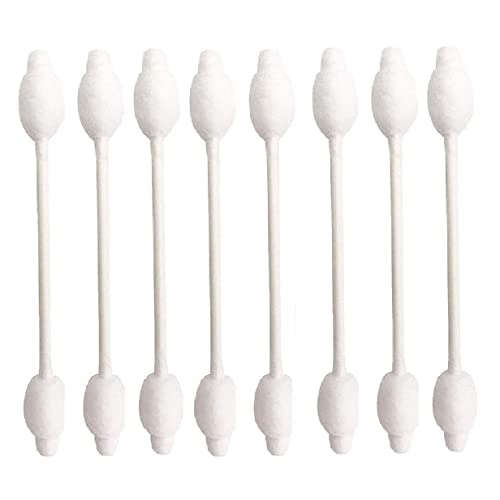} &
\includegraphics[width=1.5cm, height=2cm, keepaspectratio]{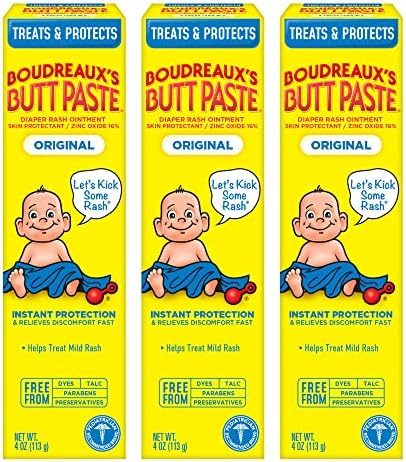} &
\includegraphics[width=1.5cm, height=2cm, keepaspectratio]{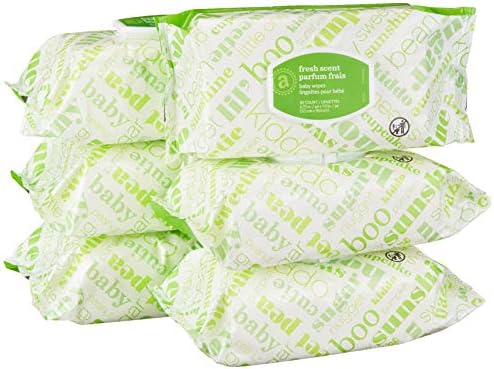} &
\includegraphics[width=1.5cm, height=2cm, keepaspectratio]{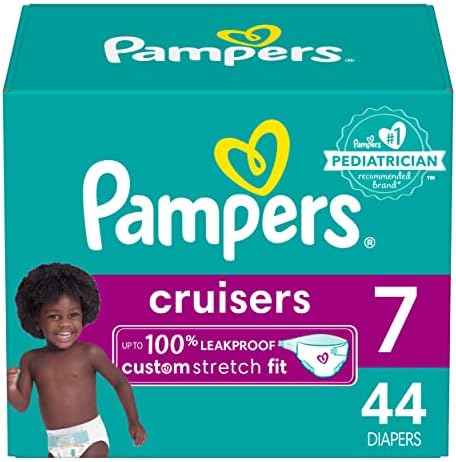} &
\includegraphics[width=1.5cm, height=2cm, keepaspectratio]{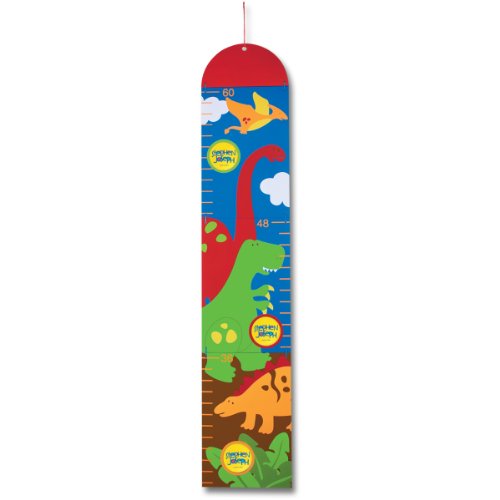} \\
\hline
\textbf{Title} & Munchkin® Miracle® 360 Toddler Sippy Cup & Safety 1st Child Proof Clear View Stove Knob Covers & Baby Cotton Swabs& Boudreaux's Butt Paste Original Diaper Rash Cream & Amazon Elements Baby Wipes & Pampers Cruisers Diapers Size 7 44 Count & Stephen Joseph Growth Chart, Dino\\
\hline
\textbf{Keywords by MMSRARec} & Munchkin Miracle 360 Sippy Cup, Pink/Purple, 10 Oz, 2 Count  & Child Proof Clear View Stove Knob, Set of 5, Children Safety  &  Baby Cotton Swabs, Organic Fragrance and Chlorine-Free, 100\% Biodegradable, 4 Packs of 56  & Boudreaux's Butt Paste Original Diaper Rash Cream, Ointment for Baby, 4 oz Tube, 3 Pack  & Amazon Elements Baby Wipes, Fresh Scent, 480 Count, Flip-Top Packs, cucumber, aloe and green tea oil & \textcolor{red}{Pampers Cruisers Diapers},  Size 7, 44 Count, 2x stretchier & \textcolor{blue}{Stephen Joseph Growth Chart}, colorful, education  \\
\hline
\textbf{Recommend} & - & - & - & - & - & \checkmark & $\times$ \\
\hline
\end{tabular}
}
\end{table}

\subsection{Parameter Analysis}

\begin{figure*}[ht]
\vspace{-2em}
    \centering
    \subfloat[Analysis of $n$]{\includegraphics[width=0.49\textwidth]{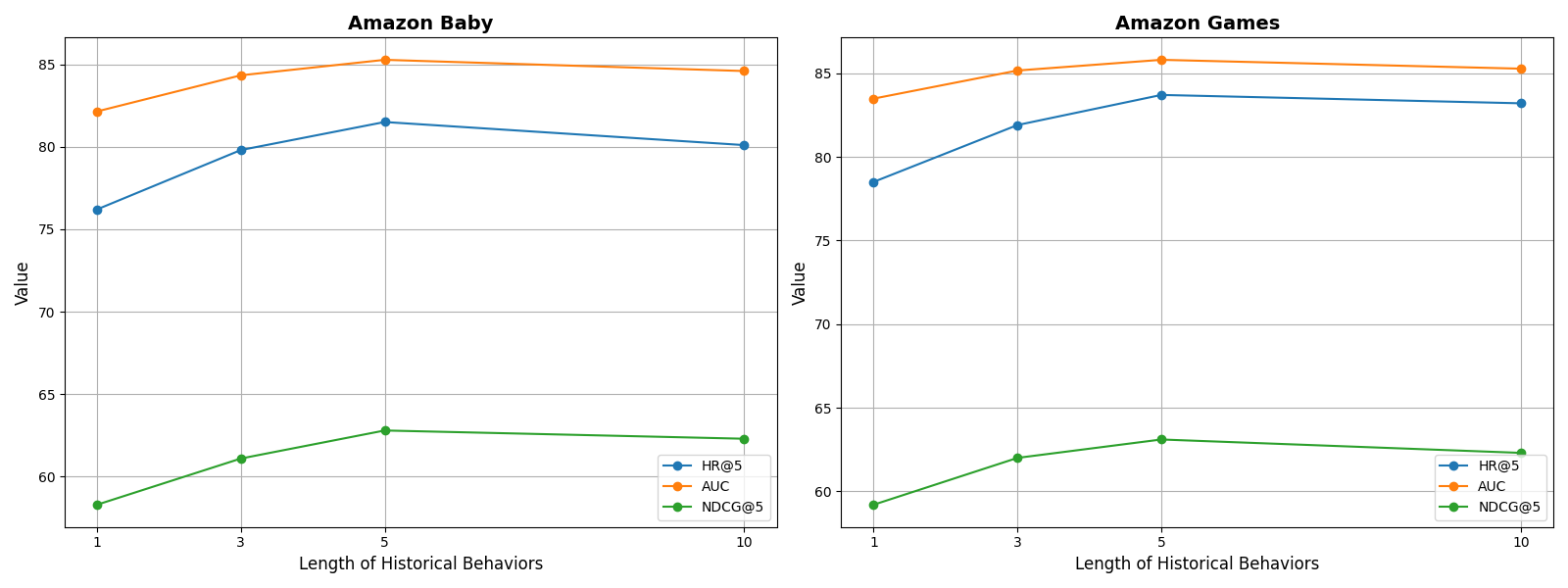}}
    \subfloat[Analysis of $|\mathcal{S}_u|$]{\includegraphics[width=0.49\textwidth]{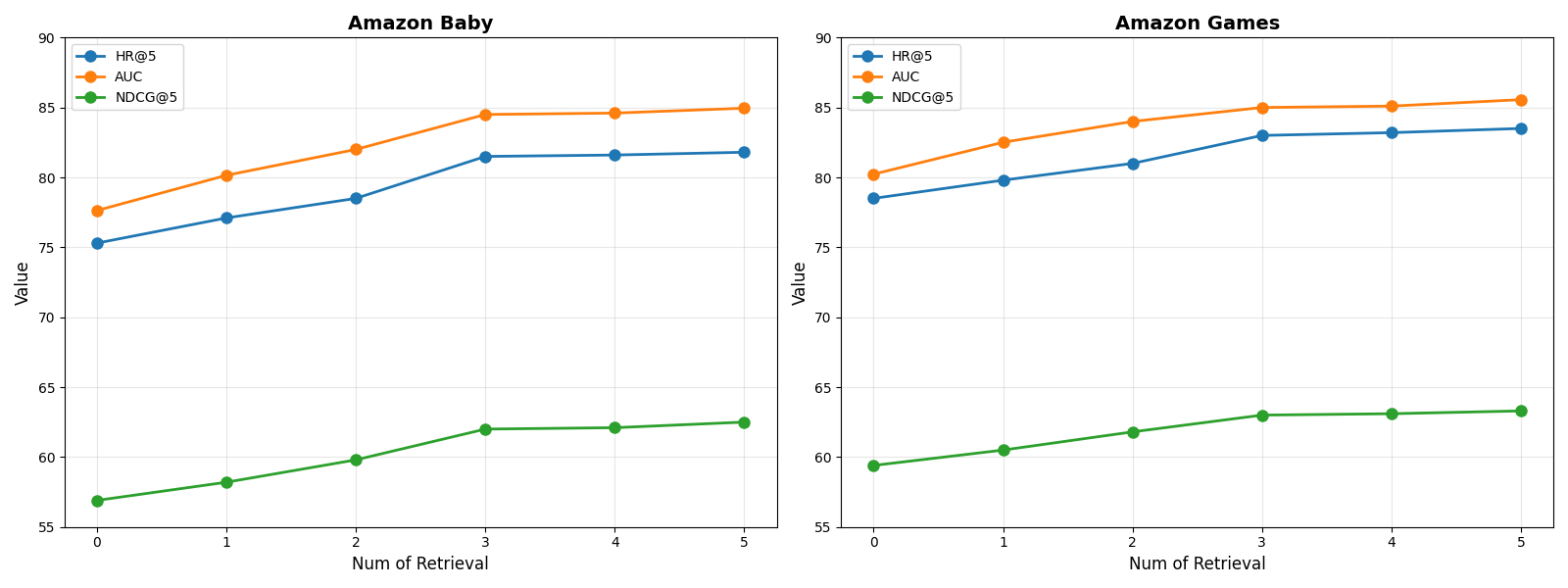}}
    \caption{Performance of adjusting the length of the user's historical behavior sequence $n$ and the number of similar users retrieved $|\mathcal{S}_u|$.  }
    \label{fig:paramter}
    \vspace{-2em}
\end{figure*}

We further investigate the impact of two key hyperparameters in the experiments on recommendation performance: the length of the user’s historical behavior sequence $n$ and the number of retrieved similar users $\mathcal{S}_u$, as shown in Fig. \ref{fig:paramter}.
When the behavior sequence is short, the limited contextual information hinders the model’s ability to accurately capture user interests.   As the sequence length increases, the model gains access to more comprehensive user information, leading to improved recommendation performance.   However, long behavior sequences may introduce noise from earlier interactions, which can interfere with the model’s judgment and prevent further performance gains. The incorporation of similar user retrieval yields a noticeable performance improvement, underscoring the importance of integrating collaborative signals into MLLM-based recommendation systems.   Similarly, retrieving too many similar users increases the context length, which can dilute the model’s focus and cause performance to converge.

In summary, achieving optimal recommendation performance requires appropriate selection of both historical behavior sequence length and the number of similar users retrieved.

\subsection{Ablation Study}

Table \ref{tab:ablation} presents the results of ablation study conducted on different stages of the model.   "w/o s1" denotes the removal of the summarization stage, where original item descriptions are used directly.   "w/o s1-training" indicates that no RLVR training is performed in the first stage.   "w/o s2" refers to the absence of similar user retrieval in the second stage. "with s2-GRU4Rec" means swith the embedding model of similar users to GRU4Rec \cite{hidasi2015session}.   "w/o s3" signifies that no training is applied in the third stage, with only the pre-trained MLLM being utilized.   "w/o s3-multitasks" represents that in the third stage, only basic instructions are used for instruction fine-tuning instead of multi-task learning.   The experimental results demonstrate that in the multimodal summarization stage, our designed reward—which comprehensively considers information loss, reconstruction difficulty, and summary length—enables the model to adaptively adjust its summarization strategy, yielding outputs that are more precise and concise compared to the original item descriptions.   During the similar user retrieval stage, the incorporation of collaborative signals assists the model in making references and judgments.   In the multi-task learning stage, our designed training tasks facilitate the model’s comprehensive understanding of the multimodal sequential recommendation task. 

\begin{table}[ht]
\vspace{-2em}
\centering
\caption{Performance of MMSRARec with different ablation strategy.}

\begin{tabular}{l|ccc|ccc}
\hline
Model & \multicolumn{3}{c|}{\textbf{Amazon Baby}} & \multicolumn{3}{c}{\textbf{Amazon Games}} \\
 & HR@5 & AUC & NDCG@5 & HR@5 & AUC & NDCG@5 \\
\hline
\textbf{MMSRARec} & \textbf{81.50} & \textbf{85.27} & \textbf{62.82} & \textbf{83.74} & \textbf{85.81} & \textbf{63.14} \\
\hline
w/o s1 & 78.91 & 83.59 & 60.12 &80.54 & 84.22 & 60.88 \\
w/o s1-training  & 79.26 &83.82 &60.58 &81.00 &84.58& 61.26\\
w/o s2  &  75.36 & 77.63 & 56.92&78.50 &80.26& 59.49\\
with s2-GRU4Rec  &  76.88 & 81.25 & 59.73&77.21 &83.36& 60.10\\
w/o s3-training  & 26.87 & 18.62 & 19.59 & 37.74 & 42.11 & 24.85 \\
w/o s3-multitasks  & 76.24& 80.18 &57.52& 77.55&81.33& 58.14\\
\hline
\end{tabular}
\label{tab:ablation}
\end{table}

\subsection{Efficiency Analysis}

Furthermore, we analyze the impact of different MLLM backbones on recommendation efficiency and compare our approach with MLLM-MSR \cite{ye2025harnessing}, a baseline method that also employs MLLMs as recommenders. The multiple inference steps and substantial time overhead of MLLM-MSR highlight the limitations of existing methods in processing user behavior sequences. Among the three compared models, Qwen2.5VL \cite{bai2025qwen2} achieves the best recommendation performance, followed by InternVL3 \cite{zhu2025internvl3}. LLaVA \cite{liu2023visual} performs the worst, which may be attributed to the lack of a training corpus relevant to multimodal sequential recommendation in its pre-training data. Additionally, while the larger Qwen2.5VL-32B model shows marginal improvement over its smaller 7B counterpart, its significantly higher training and inference costs make it unsuitable for practical online deployment. MMSRARec can accommodate arbitrary MLLM backbones, demonstrating the robustness of its architectural design.

\begin{table}[ht]
\centering
\caption{Efficiency of MMSRARec with different MLLM backbone and baseline.}
\resizebox{\textwidth}{!}{
\begin{tabular}{l|ccc|ccc|c|c}
\hline
Model & \multicolumn{3}{c|}{\textbf{Amazon Baby}} & \multicolumn{3}{c}{\textbf{Amazon Games}} & \textbf{\# Inferences} & \textbf{Time (s)} \\
 & HR@5 & AUC & NDCG@5 & HR@5 & AUC & NDCG@5 \\
\hline
MLLM-MSR-LlaVA1.5-7B \cite{ye2025harnessing}    & 73.92 & 84.39 & 58.57 & 76.39 & 85.69 & 63.73 & 6 & 7.35\\
\hline
MMSRARec-LlaVA1.5-7B  & 76.26 & 81.33 & 57.69 & 71.23 & 62.34 & 60.46 & 1 & 0.54 \\
MMSRARec-InternVL3-8B  & 78.67 & 86.12 & 60.50 & 82.96 & 85.26 & 62.33 & 1 & 0.82\\
MMSRARec-Qwen2.5VL-7B & 81.50 & 85.27 & 62.82 & \textbf{83.74} & 85.81 & 63.14 & 1 & 0.73 \\
MMSRARec-Qwen2.5VL-32B  & \textbf{82.05} & \textbf{87.28} & \textbf{66.75} & 82.56 & \textbf{87.41} & \textbf{65.88} & 1 & 1.13\\
\hline
\end{tabular}
}
\label{tab:backbone}
\vspace{-1em}
\end{table}

\section{Conclusion and Future Work}

This paper proposes a novel method named MMSRARec for multimodal sequential recommendation, based on a multimodal large language model (MLLM).    We first employ the MLLM to summarize items into concise keywords and fine-tune the model using rewards that incorporate summary length, information loss, and reconstruction difficulty, thereby enabling adaptive adjustment of the summarization strategy.    Inspired by retrieval-augmented generation, we convert collaborative signals into corresponding keywords and integrate them as contextual input.    Finally, we apply supervised fine-tuning with multi-task learning to align the MLLM’s capabilities with the requirements of multimodal sequential recommendation.    This approach leverages the MLLM’s strong semantic understanding and in-context learning abilities, introduces previously overlooked collaborative signals into the MLLM, and reduces computational and time costs through once MLLM inference.    Experiments on three real-world recommendation datasets demonstrate that MMSRARec achieves a better understanding of multimodal item information and user preferences, enabling accurate and interpretable recommendations.

Currently, the MLLM backbone used in MMSRARec contains 7B parameters, leading to longer inference times compared to traditional sequential recommendation models.    In the future, we plan to explore methods such as distillation, output decoding, and pruning to replace the current backbone with a smaller model, thereby accelerating MMSRARec without compromising performance.

%
%
%
\bibliographystyle{splncs04}
\bibliography{myref}
\end{document}